\documentclass[aps,twocolumn]{revtex4}
\usepackage{newlfont}
\usepackage{amssymb}
\usepackage{amsfonts}
\usepackage{amsmath}
\usepackage{graphicx}
\usepackage{bm}

\usepackage{graphicx}
\usepackage{epsfig}
\usepackage{newlfont}
\usepackage{amssymb}
\usepackage{amsfonts}
\usepackage{amsmath}
\usepackage{graphicx}
\usepackage{bm}

\begin{document}





\title{Bound Genuine Multisite Entanglement: Detector of Gapless-Gapped Quantum Transitions in Frustrated Systems}


\author{Aditi Sen(De) and Ujjwal Sen}

\affiliation{Harish-Chandra Research Institute, Chhatnag Road, Jhunsi, Allahabad 211 019, India}

\begin{abstract}
  
We define a multiparty entanglement measure, called generalized geometric measure, that can detect and quantify genuine multiparty entanglement for any number of parties. The quantum phase transitions in exactly solvable models like the anisotropic XY model can be detected by this measure. We find that 
the multisite measure can be a useful  tool to detect quantum phenomena in 
more complex systems like quasi 2D and 2D frustrated Heisenberg antiferromagnets.
We propose an order parameter, called bound generalized geometric measure, in the spirit of bound 
quantum states, that can recognize the gapless and gapped phases of the frustrated models
by its sign. The vanishing of the order parameter therefore  signals the transition between such phases.


\end{abstract}

\maketitle

\section{Introduction and Main Results}

 The rapid development of the theory of entanglement over the last decade or so \cite{HHHH-RMP}, and its usefulness in communication systems and 
 computational devices, as well as the experimental observations of entangled states in a variety of distinct physical systems \cite{ref-boi},
 have attracted a lot of attention 
 from different branches of physics, including condensed matter and ultra-cold gases 
\cite{ref-reviews1, ref-reviews2}.
It has been argued 
that entanglement can be used as a ``universal detector'' of quantum phase transitions,
with most of the studies being on the 
behavior of 
\emph{bipartite} 
entanglement
\cite{Wootters, logneg}.
%
A more natural way  to study the 
many-body systems would be to consider multipartite entanglement, as almost all naturally occurring multisite quantum states are 
genuinely multi-party entangled.  Such an enterprise is however limited by the intricate nature of 
entanglement theory  
in the multisite scenario. In particular, 
only a few multisite entanglement measures are known, and moreover their computation are difficult \cite{HHHH-RMP}. 

Multipartite states can have different hierarchies according to their entanglement quality and quantity. The simplest example is for 
three-particle states, where there are fully separable, biseparable, and genuine multipartite entangled states. 
A measure of genuine multiparty entanglement, quantifies, so to say, the ``purest'' form multiparty entanglement. 
In this paper, we define an entanglement measure, called generalized geometric 
measure (GGM), that can detect and quantify \emph{genuine} multiparticle entanglement. Interestingly, the 
measure is computable for arbitrary pure states of multiparty systems in arbitrary dimensions
and arbitrary number of parties, and therefore
can turn out to be a useful tool to detect quantum many-body phenomena, like quantum phase
transitions. In this respect, GGM has the potential of gaining the same status in applications of multiparty entanglement theory, as that of logarithmic negativity \cite{logneg} in 
the bipartite domain.

As an initial testing ground,
we  
use the GGM to successfully detect quantum phase transitions in the 
anistropic XY model on a chain of spin-1/2 particles \cite{Barouch-McCoy}. 
Our main aim however is to apply the measure to states of frustrated spin systems, for which 
the phase diagrams are not exactly known.  Frustrated many-body systems are a center of interest 
in condensed matter physics due to the typically rich and novel phase diagrams in such systems.
Moreover,
experimental realizations of many metal oxides, including those exhibiting high-Tc superconductivity, typically have frustrated interactions 
in their Hamiltonians \cite{snajh-er-jor,5mlParacetamol-diyechhi}.
As paradigmatic  representatives of such systems, we 
consider (i) the quasi 2D antiferromagnetic \(J_1-J_2\) Heisenberg model
with  nearest neighbor couplings,  \(J_1\), and next-nearest neighbor couplings, \(J_2\)
\cite{Majumdar-Ghosh, White-Affleck, Mikeska}, and (ii) the 
frustrated \(J_1-J_2\)  model on a square lattice \cite{snajh-er-jor} (see Fig. 1).
\begin{figure}[h!]
\label{fig-chhobi}
\begin{center}
\epsfig{figure= 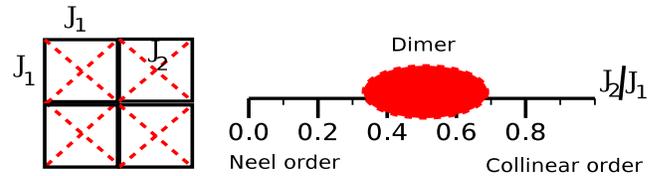, height=.15\textheight,width=0.47\textwidth}
\caption{(Color online.) Two-dimensional \(J_1-J_2\) model, with  vertical and horizontal couplings, \(J_1\), and diagonal couplings, \(J_2\) The predicted phase diagram is also schematically shown.}
\end{center}
\end{figure}

For studying such systems, we introduce an order parameter which is the difference between 
the GGM (\({\cal E}\)) and its second derivative with respect to the system parameter, \(\mu\),
 that drives 
the transitions in the system. We call the quantity as ``bound GGM'', and is given  by 
%
%
\begin{equation}
{\cal E}_B \equiv {\cal E} - \frac{d^2 {\cal E}}{d\mu^2}. \nonumber 
\end{equation}
%
%
%

%
%
The ground state manifold of the quasi 2D \(J_1-J_2\) system is not known exactly, except  at the Majumdar-Ghosh 
point \cite{Majumdar-Ghosh},  i.e. for \(\alpha = J_2/J_1 = 0.5 \), where the system is highly frustrated, and presents two dimer states as its ground states. However, 
exact diagonalization and  group theoretical studies show that the system is gapless, and hence critical, in the weakly frustrated regime, 
namely \(0 \leq \alpha \lesssim 0.24\) \cite{Majumdar-Ghosh, White-Affleck}. 
For higher coupling ratio \(\alpha\), the system enters a dimerized regime, and is gapped \cite{Majumdar-Ghosh, eita-AKLT}.  
We study the GGM for this system by exact diagonalization,
%
and  show that the bound GGM vanishes at the fluid-dimer transition point \(\alpha \approx 0.24\).
Note here that it is known that bipartite entanglement cannot 
detect the 
gapless phase
\cite{bipartite_MG} (cf. \cite{Chhajlany}). We find that the bound GGM is positive in the gapless phase while it 
becomes negative in the gapped one. The Majumdar-Ghosh point can also be detected 
by the GGM.

Finally we apply our measure of genuine multisite entanglement to the ground state of the 2D Heisenberg system.
As depicted in Fig. 1, 
the N{\' e}el and collinear ordered phases (the gapless phases dissociated 
by a phase having a finite gap between the singlet ground state and the excited states. 
We show that the bound GGM can detect both the quantum phase transitions -- from the N{\' e}el  phase to the dimerized one at \(\alpha \approx 0.38\), as well as the transition from
the dimer to the collinear phase at \(\alpha \approx 0.69\), as predicted, even for 
relatively small system-size. 
 Like in the \(J_1-J_2\) ring, the positivity (negativity) of the bound GGM indicates the gapless 
(gapped) phase.

Armed with these findings, we propose that the bound GGM can 
potentially be used for detecting gapped/gapless phases in many-body systems:
\begin{eqnarray}
\label{onek-deri-hoye-gyalo}
{\cal E}_B > 0 \Rightarrow \mbox{gapless}, \quad {\cal E}_B < 0 \Rightarrow \mbox{gapped}.
\end{eqnarray}
This leads to an analogy with the thermodynamics of bound entanglement \cite{wwwww, HHHH-RMP}. 
Analogous to the first law of thermodynamics, internal energy = free energy + work done, a thermodynamic
equation of entanglement was written: Entanglement cost = distillable entanglement + bound entanglement, 
where the bound entanglement is the amount of entanglement necessary to keep the  transition (under local 
quantum operations and 
classical communication) from becoming irreversible. As another face of this entanglement-energy analogy,
a negative value of \({\cal E}_B\), assuming the thesis in Eq. (\ref{onek-deri-hoye-gyalo}),
indicates that the system needs a nonzero amount of energy to free itself from its ground state. We hope that 
this can help us in a quantification of the first law of the emerging entanglement thermodynamics \cite{thermodynamics, HHHH-RMP}.
This is the reason for calling \({\cal E}_B\) as \emph{bound} GGM \cite{qqqqq}.


\section{Generalized Geometric Measure}

Let us begin by  defining the generalized geometric measure. As mentioned above, 
GGM will quantify the genuineness of multiparty entanglement.  
An \(N\)-party pure quantum state is said to be
genuinely \(N\)-party entangled, if it is not a product across
any bipartite partition. The simplest examples of genuine tripartite entangled states are the Greenberger-Horne-Zeilinger  \cite{GHZ} 
and W  \cite{W-state} states. The GGM of  an \(N\)-party pure quantum state \(|\psi\rangle\) is defined as
\begin{equation}
{\cal E} ( |\psi\rangle ) = 1 - \Lambda^2_{max} (|\psi\rangle ), 
\end{equation}
where  \(\Lambda_{max} (|\psi\rangle ) =
\max | \langle \phi|\psi\rangle |\), with the maximization being over all pure states \(|\phi\rangle\)
that are not genuinely \(N\)-party entangled.   Note that the maximization performed in GGM is different from  
the maximization in the geometric measure  of Ref. \cite{GM} (cf. \cite{hierarchy}). 

\subsection{Properties}

Clearly, \({\cal E}\)
 is vanishing for all pure multiparty states that are not genuine multiparty entangled, and non-vanishing for others. We considered this quantity 
for four-party states in Ref. \cite{amadertele}, and showed it to be a mono-
tonically decreasing quantity under local quantum operations and classical communication (LOCC).
Applications of GGM to quantum many-body systems requires us to find its properties for an arbitrary number of parties.

Let \(|\psi\rangle\) be an \(N\)-party pure quantum state in the tensor product Hilbert space \({\cal H}_{A_1} \otimes {\cal H}_{A_2} \otimes \ldots \otimes {\cal H}_{A_N}\). 
Therefore, the maximization in 
\begin{equation}\Lambda_{\max}(|\psi\rangle_{A_1 A_2 \ldots A_N}) = \max_{|\phi\rangle_{A_1 A_2 \ldots A_N}} |\langle \phi | \psi \rangle|
 \end{equation}
is over all pure quantum states \(|\phi\rangle_{A_1 A_2 \ldots A_N}\), in \({\cal H}_{A_1} \otimes {\cal H}_{A_2} \otimes \ldots \otimes {\cal H}_{A_N}\), 
that are not genuinely multiparty entangled, which is a rather large class of states. 
Note however, that the square of \(\Lambda_{\max}(|\psi\rangle_{A_1 A_2 \ldots A_N})\) can 
be interpreted as the Born probability of some outcome in a quantum measurement on the state \(|\psi\rangle\). 
Now, entangled measurements cannot be worse than the product ones for any set of subsystems. Therefore, in the maximization, we do not need to consider the
\(|\phi\rangle_{A_1 A_2 \ldots A_N}\) that are product in a partition of \(A_1, A_2, \ldots, A_N\) into three, four, ... sets. 
The only \(|\phi\rangle_{A_1 A_2 \ldots A_N}\) that are to be considered are the ones that are a product in a \emph{bi}-partition of \(A_1, A_2, \ldots, A_N\).
This greatly reduces the class over which the maximization is carried out. 
Let \({\cal A}: {\cal B}\) be such a bi-partition. 
Then, \(\max |\langle \phi | \psi \rangle|\), where the maximization is carried over the \(|\phi\rangle\) that are product across \({\cal A}: {\cal B}\), 
is the maximal Schmidt coefficient, \(\lambda_{{\cal A}: {\cal B}}\), of the state \(|\psi\rangle_{A_1 A_2 \ldots A_N}\) in the \({\cal A}: {\cal B}\) bipartite split.
\(\Lambda_{\max}(|\psi\rangle_{A_1 A_2 \ldots A_N})\) is therefore the maximum of all such maximal Schmidt coefficients in bipartite splits. 
Note that the 
\(\lambda\)'s 
involved 
in this
closed form for \(\Lambda_{max}\)
are all increasing under LOCC \cite{Vidal-Nielsen}. We have therefore proven the 
following theorem.\\
\noindent \textbf{Theorem.} 
\emph{The generalized geometric measure of  \(|\psi\rangle_{A_1 A_2 \ldots A_N}\) is given by 
%
\begin{equation}
{\cal E}(|\psi\rangle) = 1 - \max \{\lambda^2_{{\cal A}: {\cal B}} | {\cal A} \cup {\cal B} = \{1,2,\ldots, N\}, {\cal A} \cap  {\cal B} = \emptyset\}.
 \end{equation}
It is computable for a multiparty pure state of an arbitrary number of parties, and of arbitrary dimensions. Also, it is monotonically decreasing under LOCC.
}

\section{Anisotropic  XY model}

The one-dimensional XY model with \(N\) lattice sites is described by the Hamiltonian
\begin{equation}
\label{eq_XY_H}
 H_{XY} = \frac{J}{2} \left(\sum_{i=1}^{N} (1 + \gamma) \sigma^x_i \sigma^x_{i+1} + (1 - \gamma) \sigma^y_i \sigma^y_{i+1}\right) + h \sum_{i=1}^{N} \sigma_i^z, 
\end{equation}
where \(J\) is the coupling constant,  \(\gamma \in [0,1] \) is the anistropy parameter, \(\sigma\)'s are the Pauli matrices, 
and \(h\) represents the magnetic field in the transverse direction. The quantum transverse Ising and the transverse XX models correspond to two 
extreme values of \(\gamma\), which are resepectively \(\gamma =1\) and \(\gamma =0\). 
This model can be diagonalized  by the Jordan-Wigner transformation \cite{Barouch-McCoy}. Apart from its other interests, it is the simplest 
model which shows a \emph{quantum} phase transition, driven by the magnetic field,
at zero temperature. 
It is known to be detectable by using bipartite entanglement measures \cite{fazio-Nielsen}, 
like concurrence
\cite{Wootters}. However, evaluating GGM  will additionally  quantify the nature of genuine multiparty entanglement of the ground state in this model, especially
as it crosses the transition point.

The diagonalization of this model can be achieved by 
introducing the Majorana fermions
\begin{equation}
c_{2l -1} = (\Pi_{i=1}^{l-1} \sigma_i^z)\sigma^x_l; \quad  c_{2l} = (\Pi_{i=1}^{l-1} \sigma_i^z)\sigma^y_l. 
\end{equation}
The  Hamiltonian in Eq. (\ref{eq_XY_H}) thereby reduces to a quadratic fermionic Hamiltonian \cite{Barouch-McCoy}.
The eigenvalues of the reduced density matrix of \(L\) sites of the ground state of this system
 can be obtained by using the above formalism \cite{ref-reviews2}, and is given by 
\begin{equation}
 e_{x_1 x_2 \ldots x_l} = \prod_{i=1}^{L} \frac{1 + (-1)^{x_i} \nu_i}{2}, \quad x_i = 0, 1 \phantom{,} \forall i, 
\end{equation}
 where \(\nu_i\)'s are the eigenvalues of \(G_L\), which in turn is given by
 \(B_L = G_L \otimes \left[\begin{array}{cc}
0 & 1  \\
-1 & 0 \\
\end{array}\right]\), with 
\begin{equation}
\label{beRal-er-talobya-sho}
G_L=\left[
\begin{array}{cccc}
g_0 &  \cdot & \cdot & g_{L-1} \\
\cdot &  \cdot & \cdot & \cdot\\
- g_{L-1} & \cdot & \cdot & g_{0} \\
\end{array}\right], 
%
%
%
%
%
%
 B_L =  \left[
\begin{array}{cccc}
\Pi_0 & \cdot & \cdot & \Pi_{L-1} \\
\cdot & \cdot & \cdot & \cdot\\
- \Pi_{L-1} & \cdot & \cdot & \Pi_{0} \\
\end{array}\right]. \nonumber
\end{equation}
Here, \(\Pi_l = \left[
\begin{array}{cc}
0 & g_l  \\
-g_{-l} & 0 \\
\end{array}\right]\), and the real coefficients, \(g_l\), are given by
\begin{equation}
 g_l = \frac{1}{2 \pi} \int_{0}^{2 \pi} d\phi e^{- i l \phi} \frac{\cos \phi - 
\lambda - i \gamma \sin \phi}{|\cos \phi - \lambda - i \gamma \sin \phi|}, 
\end{equation}
where \(\lambda = J/h\).

The derivative of GGM  of the ground state, for different anistropy parameters \(\gamma\), clearly 
shows a logarithmic divergence at the transverse field given by \(\lambda =1\), as seen in Fig. 2.   
Note also  that the ground state 
of the transverse Ising model (\(\gamma =1\)) has higher genuine multipartite entanglement as  compared to the ground states for other values of \(\gamma\).
This result may help us to understand the success of the dynamical states of the transverse Ising model as a substrate for efficient quantum computation \cite{raus-briegel}.

\vspace{0.7cm}

\begin{figure}[h!]
\label{fig-chhobi-ek}
\begin{center}
\epsfig{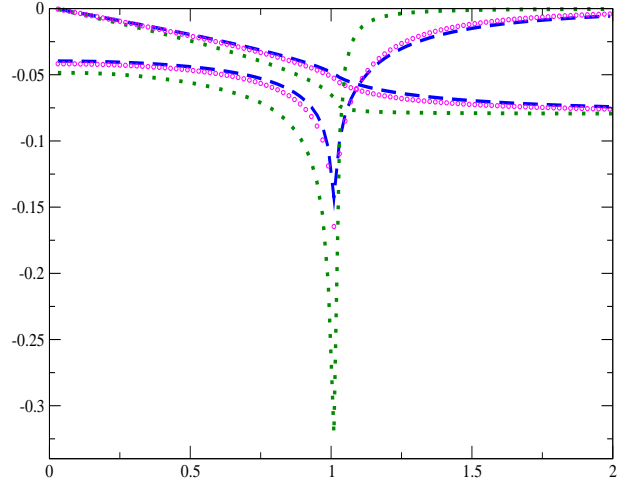}
\caption{(Color online.) GGM of the transverse XY model. 
The GGM (actually \({\cal E} - 1/2\)) and its derivative  (both dimensionless)
are plotted on the vertical axis for the anistropic transverse XY model 
for different anisotropy parameters \(\gamma\), against the dimensionless system parameter \(\lambda\) on the horizontal axis.  
The dashed (blue), circled (pink), and dotted (green) lines are respectively for the Ising (\(\gamma =1\)), \(\gamma = 0.8\), and \(\gamma = 0.2\) models.  
The derivatives of GGM  diverges at the quantum critical point \(\lambda = 1\). 
}
\end{center}
\end{figure}

\section{Quasi 2D Frustrated \(J_1-J_2\) Model}

We will now consider the frustrated  quasi two-dimensional \(J_1-J_2\) Heisenberg model,
in the case when both the nearest neighbor couplings, \(J_1\), and 
the next-nearest neighbor couplings, \(J_2\), are antiferromagnetic. Apart from its other interests, 
the intense interest for studying this model lies in the fact that it is similar to real systems, like \(\mbox{SrCuO}_{2}\) \cite{experiment}.
The Hamiltonian of this model, with \(N\) lattice sites on a chain, is
\begin{equation}
 H_{1D} = J_1 \sum_{i=1}^{N} \vec{\sigma}_i \cdot \vec{\sigma}_{i+1} + J_2 \sum_{i=1}^N \vec{\sigma}_i \cdot \vec{\sigma}_{i+2},
\end{equation}
where \(J_1\) and \(J_2\) are both positive, and where periodic boundary condition in assumed.
The ground state  and the energy gap of this  model were studied 
by using exact diagonalization, density matrix renormalization group method,
bosonization technique, etc \cite{Mikeska}. For an even number of sites, the ground state at 
the Majumdar-Ghosh point (\(\alpha = J_2/J_1 = 0.5\)),  is doubly degenerate, and the ground state manifold is spanned by the two dimers
\(|\psi_{MG}^{\pm}\rangle = \Pi_{i=1}^{N/2} (|0\rangle_{2i} |1\rangle_{2i \pm 1} -  |1\rangle_{2i} |0\rangle_{2i \pm 1})\), 
and the model is gapped at this point \cite{Majumdar-Ghosh}. 
For \(\alpha=0\), the Hamiltonian reduces to the \(s=1/2\) Heisenberg antiferromagnet and hence the the ground state, which is a spin fluid state having  
gapless excitations \cite{Griffiths-Yang}, can be obtained by Bethe ansatz. It is known that  at \(\alpha \approx 0.2411\), a phase transtion from fluid to  dimerization occurs \cite{gap_transition}.

The genuine multipartite entanglement measure clearly signals the Majumdar-Ghosh point (See Fig. 3). The fluid-dimer 
tranition at \(\alpha \approx 0.24\) can also be detected by the vanishing of the bound GGM as the order parameter (for \(\mu = \alpha\)) (see Fig. 3). 
%
Moreover, 
 \({\cal E}_B >0\) signals the gapless phase, while  \({\cal E}_B <0\) indicates the gapped phase.

 \vspace{0.7cm}
\begin{figure}[h!]
\label{fig-chhobi-dui}
\begin{center}
\epsfig{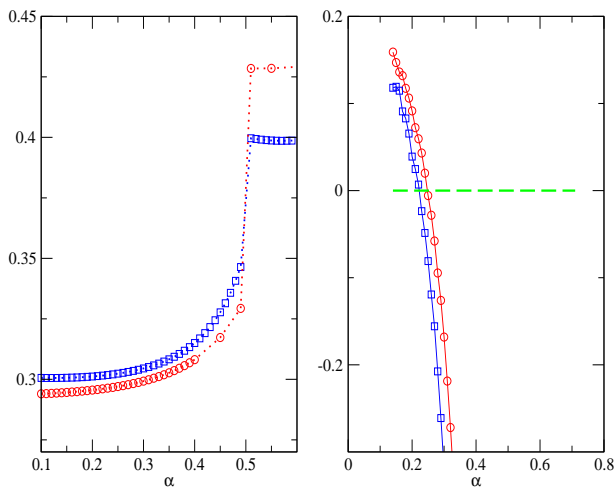}
\caption{
(Color online.) GGM and bound GGM for the quasi-2D frustrated antiferromagnet. 
The left figure is for the GGM on the vertical axis
 against \(\alpha\) on the horizontal. 
The Majumdar-Ghosh point at \(\alpha = 0.5\) 
is clearly signaled. 
The figure on the right is for the bound GGM on the vertical axis, against \(\alpha\) on the horizontal, 
and the fluid-dimer transition is signaled by the vanishing of the bound GGM. 
 The two curves are for \(8\) (red circles) and \(10\) (blue squares) spins
in both the figures.
The GGM, bound GGM, and \(\alpha\) are all dimensionless. 
}
\end{center}
\end{figure}

\section{2D Frustrated \(J_1-J_2\) Model}


We now consider spin-1/2 particles on a square lattice, where nearest neighbor spins (both vertical and horizontal)
 on the lattice are coupled by Heisenberg interactions, with coupling strengths \(J_1\), and where 
all diagonal spins are coupled by Heisenberg interactions, with coupling strengths \(J_2\) (see Fig. 1). 
This 2D model have attracted a lot of interest \cite{many_theory} 
due to its connection 
with the high \(T_c\)-superconductors and 
its similarity with magnetic materials 
like \(\mbox{Li}_2 \mbox{VOSiO}_4\) and  \(\mbox{Li}_{2}\mbox{VOGeO}_4\) \cite{synthesis}. 
Although the different phases of the ground state of this model is well-studied,
there seem to exist reasons to believe in further secrets hidden. 
%
The Hamiltonian of the system is therefore given by 
\begin{equation}
 H_{2D} = J_1 \sum_{\langle i,j \rangle} \vec{\sigma}_i \cdot \vec{\sigma}_{j} + 
J_2 \sum_{i,j \in {\cal D}} \vec{\sigma}_i \cdot \vec{\sigma}_{j}.
\end{equation}
Both \(J_1\) and \(J_2\) are antiferromagnetic (\(>0\)).

In the classical limit, the model exhibits only a first-order phase transition from N{\' e}el to collinear at \(\alpha = J_2/J_1 = 0.5\).
The phase diagram changes its nature, when  quantum fluctuations  are present, and in this case, the exact phase boundaries are not known. 
It is expected that two long range ordered (LRO) ground state phases are separated by quantum paramagnetic phases without LRO. 
Different methods, like exact diagonalization, series expansion methods, field-theory methods \cite{Richter10-Kim-Singh}, etc., 
applied to this model, predict that the first transition from N{\' e}el to dimer accurs at \(\alpha \approx 0.38\) while other one happens at \(\alpha \approx 0.66\). 
%
Recent experimental observations and proposals of detecting such phases in the laboratory demand the precise quantification of the phase diagram of this model at low temperature. Towards this aim, we show that even for  
relatively small system size, the order parameter based on the genuine multipartite entanglement measure 
(the \({\cal E}_B\), introduced above) can detect and quantify the phase diagram accurately.

We perform exact diagonalization to find the ground state of the model, and we show that 
both the transitions -- N{\' e}el to dimer and dimer to collinear can be signaled by the bound GGM. 
A synopsis of these facts is given in Fig. 4. Precisely, we have found that \({\cal E}_B\) vanishes 
at \(\alpha \approx 0.38\) and again at \(\alpha \approx 0.69\). As observed in the case of the 
quasi 2D \(J_1-J_2\) model, \({\cal E}_B\) is positive in the gapless phases while it is 
negative in the intermediate gapped phase. 

 \vspace{0.7cm}
\begin{figure}[h!]
\label{fig-chhobi-tin}
\begin{center}
\epsfig{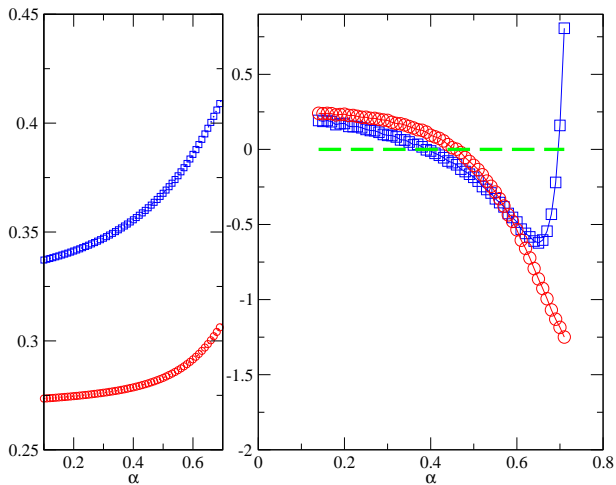}
\caption{
(Color online.) GGM and bound GGM for the 2D frustrated antiferromagnet. The horizontal axes in both the figures represent \(\alpha\). 
The left figure is for the GGM on the vertical axis
while the right one is for the bound GGM. 
The vanishing of the bound GGM signals both the N{\' e}el-to-dimer and the dimer-to-collinear transitions. 
And the gapped (gapless) phase(s) is (are) signaled by a negative (positive) bound GGM.
The two curves are for \(9\) spins on a \(3 \times 3\) square lattice (red circles), and \(12\) spins 
on a \(3 \times 4\) rectangular lattice (blue squares)
in both the figures.
The GGM, 
bound GGM, and \(\alpha\) are all dimensionless. 
}
\end{center}
\end{figure}

\section{Conclusions}

Multipartite entangled states can be classified according to their separability in different partitions. Due to the complex classification, it is hard to obtain a unique multipartite entanglement measure. Instead of quantifying all the classes of multipartite states, we define an entanglement measure, called generlized geometric measure, that quantifies the ``purest'' form of multiparty entanglement, the genuine multipartite entanglement. This is akin to the situation in bipartite pure states, where there is essentially a unique entanglement measure, while
mixed bipartite states allows a number of such measures \cite{HHHH-RMP}. In the case of multiparty states, we find ``pure'' and ``non-pure'' forms of entanglement, even within the class of pure states, where the ``pure'' part can be 
quantified by the generalized geometric measure defined here. Moreover, we found that the measure can be reduced to a simplified closed form, and hence is computable for arbitrary dimensions and arbitrary number of parties.


We then applied this measure to detect  phase diagrams in quantum many-body systems. After successfully verifying that the measure can detect quantum fluctuation driven phase transitions in the exactly solvable models like the XY Hamiltonian, we applied the generalized geometric measure to frustrated models like quasi two dimensional and two dimensional antiferromagnetic \(J_1-J_2\) models. 
In the latter case, the phase diagram is not known exactly, although there has been several predictions by different methods. In this paper, we show that an order parameter, called bound GGM, based on the multi-site entanglement 
measure defined, can signal the phase boundaries in both the models. Moreover, we found the the order parameter is positive when the system is gapless and negative in the gapped phase. We propose that the sign of the 
bound GGM can indicate whether a many-body system is gapped or gapless, and point to its implication 
for the first law of entanglement thermodynamics.


\acknowledgments

We acknowledge partial support from the Spanish MEC (TOQATA (FIS2008-00784)).


\vspace{1cm}


\begin{thebibliography}{99}

\bibitem{HHHH-RMP} R. Horodecki,
 P. Horodecki,  M. Horodecki, and K. Horodecki,
 Rev. Mod. Phys. \textbf{81}, 865 (2009).


\bibitem{ref-boi} M.A. Nielsen and I.L. Chuang,  \emph{Quantum Computing and Quantum Information} (Cambridge University Press,
Cambridge, 2000); \emph{Lectures on Quantum Information} eds. D. Bru{\ss} and G. Leuchs (Wiley, Weinheim, 2006).


\bibitem{ref-reviews1}  M. Lewenstein,
 A. Sanpera, V. Ahufinger, B. Damski, A.  Sen(De), and U. Sen,
Adv. Phys. \textbf{56}, 243 (2007).

\bibitem{ref-reviews2} L. Amico,
R. Fazio, A. Osterloh, and V. Vedral,
Rev. Mod. Phys. \textbf{80}, 517 (2008).


\bibitem{Wootters} W.K. Wootters, Phys. Rev. Lett. \textbf{80}, 2245 (1998).

\bibitem{logneg} G. Vidal and R.F. Werner, Phys. Rev. A \textbf{65}, 032314 (2002).











\bibitem{Barouch-McCoy} E. Lieb, T. Schultz, and D. Mattis, Ann. Phys. \textbf{16}, 407 (1961);
E. Barouch, B.M. McCoy, and M. Dresden, Phys. Rev. \textbf{2}, 1075 (1970);
E. Barouch and B.M. McCoy, Phys. Rev. \textbf{3}, 786 (1971).


\bibitem{snajh-er-jor} G. Misguich and C. Lhuillier, 
in \emph{Frustrated
Spin Systems}, ed. H.T. Diep (World Scientific, Singapore,
  2003);
C. Lhuillier, arXiv:cond-mat/0502464; F. Alet, A.M. Walczak, M.P.A. Fisher, Physica  (Amsterdam) \textbf{369A}, 122 (2006).




\bibitem{5mlParacetamol-diyechhi} M. Rasolt and Z. Tesanovi{\' c}, Rev. Mod. Phys. \textbf{64}, 709 (1992); M. Sigrist and T.M. Rice, \emph{ibid.} \textbf{67}, 503 (1995).

\bibitem{Majumdar-Ghosh}C.K. Majumdar and D.K. Ghosh, J. Math. Phys. \textbf{10},
1388 (1969); C.K. Majumdar and D.K. Ghosh,  \emph{ibid.} \textbf{10}, 1399 (1969).

\bibitem{White-Affleck} S.R. White and I. Affleck, Phys. Rev. B \textbf{54}, 9862 (1996).

\bibitem{Mikeska} H.J. Mikeska and A.K. Kolezhuk, in \emph{Quantum Magnetism}, 
ed. U. Schollw{\" o}ck, J. Richter, D.J.J. Farnell, and R.F. Bishop, 
(Springer, Berlin, 2004).





\bibitem{eita-AKLT} I. Affleck, T. Kennedy, E. Lieb, and H. Tasaki, Commun. Math. Phys. \textbf{115}, 477 (1988).  

\bibitem{bipartite_MG} X.-F. Qian, T. Shi, Y. Li, Z. Song, and C.-P. Sun, Phys. Rev. A \textbf{72}, 012333 (2005).
\bibitem{Chhajlany} R.W. Chhajlany, P. Tomczak, A. W{\"o}jcik, and J. Richter,  Phys. Rev. A \textbf{75}, 032340 (2007).


\bibitem{wwwww} M. Horodecki, P. Horodecki, R. Horodecki, Phys. Rev. Lett. \textbf{80}, 5239 (1998).

\bibitem{thermodynamics} P. Horodecki, R. Horodecki, and M. Horodecki, Acta Phys. Slov. \textbf{48}, 141 (1998); R. Horodecki, M. Horodecki, and P. Horodecki, Phys. Rev. A \textbf{63}, 022310 (2001). 

\bibitem{qqqqq} Note also that the equation \(\frac{d^2 \psi}{dx^2} = k^2 \psi\), for positive \(k^2\), 
is the time-independent Schr{\" o}dinger equation for \emph{bound} states of one-dimensional potential well systems
in the zero-potential region.









\bibitem{GHZ} D.M. Greenberger, M.A. Horne, and A. Zeilinger, in \emph{Bell's
 Theorem, Quantum Theory, and Conceptions of the Universe}, ed. M. Kafatos
 (Kluwer, Dordrecht, 1989).



 \bibitem{W-state} A. Zeilinger, M.A. Horne, and D. M. Greenberger, in
\emph{Squeezed States and Quantum
Uncertainty}, eds. D. Han, Y.S. Kim, and W.W.
Zachary (NASA Conference Publication 3135,
NASA, College Park, 1992);
W. D{\"u}r, G. Vidal, and J.I. Cirac, Phys. Rev. A \textbf{62},
062314 (2000).

\bibitem{GM} A. Shimony, Ann. N.Y. Acad. Sci. \textbf{755}, 675
(1995); H. Barnum and N. Linden, J. Phys. A \textbf{34}, 6787 (2001);
 T.-C. Wei and P.M. Goldbart,
Phys. Rev. A \textbf{68}, 042307 (2003).

\bibitem{hierarchy} M. Balsone, F. Dell'Anno, S. De Siena and F. Illuminatti, Phys. Rev. A \textbf{77}, 062304 (2008). 

\bibitem{amadertele} A. Sen(De) and U. Sen, Phys. Rev. A \textbf{81}, 012308 (2010).


\bibitem{Vidal-Nielsen}  M.A. Nielsen,
Phys. Rev. Lett. \textbf{83}, 436 (1999); G. Vidal,
\emph{ibid.} \textbf{83}, 1046 (1999); 
G. Vidal, 
J. Mod. Opt. \textbf{47},  355 (2000).  




\bibitem{fazio-Nielsen} A. Osterloh, L. Amico, G. Falci, and R. Fazio, Nature \textbf{416}, 618 (2002); 
T.J. Osborne and M.A. Nielsen, Phys. Rev. A \textbf{66}, 032110 (2002).

\bibitem{raus-briegel} H. J. Briegel and R. Raussendorf, Phys. Rev. Lett. \textbf{86}, 910 (2001);
R. Raussendorf and H.J. Briegel, \emph{ibid.} \textbf{86}, 5188 (2001).


\bibitem{experiment} M. Matsuda and K. Katsumata, J. Magn. Magn. Mater. \textbf{140-145}, 1671 (1995).


\bibitem{Griffiths-Yang} R.B. Griffiths, Phys. Rev. \textbf{133}, A768 (1964); C.N. Yang and C.P. Yang, \emph{ibid.} \textbf{150}, 327 (1966).

\bibitem{gap_transition} F.D.M. Haldane, Phys. Rev. B \textbf{25}, 4925 (1982); K. Okamoto and K. Nomura, Phys. Lett. A \textbf{169}, 433 (1992). 

\bibitem{many_theory}  P. Chandra and B. Doucot, Phys. Rev. B \textbf{38}, 9335 (1988);
 E. Dagotto and A. Moreo, Phys. Rev. Lett. \textbf{63}, 2148 (1989);
F. Figueirido, A. Karlhede, S. Kivelson, S. Sondhi, M. Rocek, and D. S. Rokhsar, Phys. Rev. B \textbf{41}, 4619 (1990); N. Read and S. Sachdev, Phys. Rev. Lett. \textbf{66}, 1773 (1991); H.J. Schulz and T. A. L. Ziman, Europhys. Lett. \textbf{18}, 355 (1992); R. Darradi, O. Derzhko, R. Zinke, J. Schulenburg, S.E. Kr{\"u}ger, and J. Richter, Phys. Rev. B \textbf{78}, 214415 (2008), and references therein. 


\bibitem{synthesis} R. Melzi, P. Carretta, A. Lascialfari, M. Mambrini, M. Troyer, P. Millet, and F. Mila, Phys. Rev. Lett. \textbf{85}, 1318 (2000); H. Rosner, R. R. P. Singh, W. H. Zheng, J. Oitmaa, S. L. Drechsler, and W. E. Pickett, Phys. Rev. Lett. \textbf{88}, 186405 (2002);
 R. Nath, A.A. Tsirlin, H. Rosner, and C. Geibel, Phys. Rev. B \textbf{78}, 064422 (2008);  T. Yildirim, Phys. Rev. Lett. \textbf{101}, 057010 (2008); Q. Si and E. Abrahams, Phys. Rev. Lett. \textbf{101}, 076401 (2008).
 


\bibitem{Richter10-Kim-Singh} J. Richter and J. Schulenberg, Eur. Phys. J. B \textbf{73}, 117 (2010); J.K. Kim and M. Troyer, Phys. Rev. Lett. \textbf{80}, 2705 (1998); T. Pardini and R.R.P. Singh, Phys. Rev. B \textbf{79}, 094413 (2009). 




\end{thebibliography}
\end{document}